\documentclass{appolb}%
\usepackage{graphicx}
\usepackage{amsmath}
\usepackage{amsfonts}
\usepackage{amssymb}%
\providecommand{\U}[1]{\protect\rule{.1in}{.1in}}
\pdfoutput=1
\begin{document}

\title{The light scalar $K_{0}^{\ast}(700)$ in the vacuum and at nonzero
temperature\thanks{Presented at XIII Workshop on Particle Correlations and
Femtoscopy, 22-26/5/2018, Krakow.} }
\author{Francesco Giacosa$^{1,2}$\\\address{
$^1$Institute of Physics, Jan Kochanowski University, PL-25406 Kielce, Poland
$^2$Institute for Theoretical Physics, Goethe
University, D-60438 Frankfurt am Main, Germany} }
\maketitle

\begin{abstract}
There is mounting evidence toward the existence of a light scalar kaon
$\kappa\equiv$ $K_{0}^{\ast}(700)$ with quantum numbers $I(J^{P})=\frac{1}%
{2}(0^{+}).$ Here, we recall the results of an effective model with both
derivative and non-derivative terms in which only one scalar kaonic field is
present in the Lagrangian (the standard quark-antiquark
,,seed\textquotedblright\ state $K_{0}^{\ast}(1430)$): a second
\textquotedblleft companion\textquotedblright\ pole $K_{0}^{\ast}(700)$
emerges as a dynamically generated state. A related question is the role of
$K_{0}^{\ast}(700)$ at nonzero $T$: since it is the lightest scalar strange
state, one would naively expect that it is relevant for $\pi$ and $K$
multiplicities. However, a repulsion in the $\pi K$ channel with $I=3/2$
cancels its effect.

\end{abstract}



\section{Introduction}

The lightest scalar kaonic state listed in the PDG \cite{pdg} is $K_{0}^{\ast
}(700)$ (previously called $K_{0}^{\ast}(800),$ see PDG 2016 \cite{pdg2016}
and older versions). This state, sometimes called $\kappa$, still
\textquotedblleft\textit{needs confirmation\textquotedblright}, but many works
do find a pole in that energy region, see Ref. \cite{kpoles} and refs.
therein. The PDG reports at present the following result:%
\begin{equation}
\text{pole }\kappa\text{ [PDG]: }(630\text{-}730)-i(260\text{-}340)\text{
MeV,} \label{pdgpolek}%
\end{equation}
(hence, the pole width lies between $520$-$680$ MeV), while the Breit-Wigner
(BW) mass and widths are%
\begin{equation}
\text{BW [PDG]: }m_{\kappa,BW}=824\pm30\text{ MeV , }\Gamma_{\kappa,BW}%
=478\pm50\text{ MeV.}%
\end{equation}
The BW and the pole widths are compatible, but the BW mass is somewhat larger.
There is however no friction, since BW and pole masses are different
quantities which coincide only when a resonance is narrow. This is definitely
not the case for the $\kappa$, which is a very broad state with a
width-to-mass ratio larger than $0.5.$

In a certain sense, the light $\kappa$ can be regarded as the
\textquotedblleft brother\textquotedblright\ of the light $\sigma\equiv
f_{0}(500)$ meson \cite{pdg}. This state is also very broad and for a long
time it was not clear if there is a pole on the complex plane. Now its
existence is confirmed by many studies and the state is listed in the PDG, see
also the review paper \cite{pelaezrev}. The destiny of the light $\kappa$
looks somewhat similar: its final confirmation is probably just a matter of time.

Yet, a different issue is the nature of the $\kappa\equiv K_{0}^{\ast}(700)$
and the $\sigma\equiv f_{0}(500)$. According to mounting evidence, both states
are not simple quark-antiquark states, but are rather four-quark objects,
either in the form of a tetraquark nonet together with $a_{0}(980)$ and
$f_{0}(980)$ \cite{tq} or as dynamically generated molecular-like states
\cite{lowscalars}. The $\kappa$ can be then interpreted as a
diquark-antidiquark state ($[u,d][\bar{d},\bar{s}]$ ,...) and/or as $K\pi$
state (mixing among these configurations is of course possible and rather
probable to occur). If $\kappa$ is not $\bar{q}q$, where should be the scalar
strange quarkonium? According to the quark model \cite{isgur} and modern
chiral approaches \cite{elsm}, the lightest $\bar{q}q$ kaonic state
($u\overline{s}$,...) is the well-established $K_{0}^{\ast}(1430)$ (similarly,
the lightest scalar/isoscalar quarkonium is the state $f_{0}(1370)$). The
question that we review in this work is the link between the standard state
$K_{0}^{\ast}(1430)$ and the dynamically generated state $K_{0}^{\ast}(700).$
We find (see Sec. 2) that the $\pi K$ loops dressing $K_{0}^{\ast}(1430)$
generate $K_{0}^{\ast}(700)$ as a companion pole (a peculiar four-quark
object) \cite{milena} (similarly, the $a_{0}(980)$ emerges as a companion pole
of $a_{0}(1450)$ \cite{a0}).

There is however a related important question: if the light $\kappa$ is
existent, should it be included into thermal hadronic models \cite{andronic}?
At a first sight, the answer is `yes'. In fact, the light $\kappa$ is the
second-lightest state with nonzero strangeness, thus potentially relevant.
Yet, a detailed analysis of the problem \cite{bronio} shows that one should
better \textit{not} include this state into a thermal model (see Sec. 3).
Namely, also repulsive channels contribute to the thermodynamics
\cite{dashen,hiller,pok}. Just as for the $f_{0}(500)$ whose contribution is
cancelled by $\pi\pi$ scattering with $I=2$, the contribution of the $\kappa$
is cancelled by the repulsion in $\pi K$ channel with $I=3/2$ . Thus, the
easiest thing to do is to neglect both the $f_{0}(500)$ and the $K_{0}^{\ast
}(700)$ when studying hadronic thermal models for the late stage of heavy ion collisions.

\section{The light $\kappa$ in the vacuum}

As a first step, we write down a Lagrangian that contains \textit{only one}
scalar state $K_{0}^{\ast}$, to be identified with $K_{0}^{\ast}(1430),$
coupled to $K\pi$ pairs:
\begin{equation}
\mathcal{L}_{K_{0}^{\ast}}=aK_{0}^{\ast+}K^{-}\pi^{0}+bK_{0}^{\ast+}%
\partial_{\mu}K^{-}\partial^{\mu}\pi^{0}+\ldots\text{ ,} \label{lagk}%
\end{equation}
where dots refer to other isospin channels. Note, there is \textit{no}
$\kappa\equiv K_{0}^{\ast}(700)$ into the model (yet). There are both
derivative and non-derivative terms: the former naturally dominates in the
context of chiral perturbation theory and also emerge from the extended Linear
Sigma Model \cite{elsm}. The decay width reads:
\begin{equation}
\Gamma_{K_{0}^{\ast}\rightarrow K\pi}(m)=3\frac{\left\vert \vec{k}%
_{1}\right\vert }{8\pi m^{2}}\left[  a-b\frac{m^{2}-M_{K}^{2}-M_{\pi}^{2}}%
{2}\right]  ^{2}F_{\Lambda}(m)\text{ ,}%
\end{equation}
with the vertex function $F_{\Lambda}(m)=\exp(-2\vec{k}_{1}^{2}/\Lambda^{2})$.
Here, $\Lambda$ is an energy scale describing the nonlocal nature of mesons
\cite{nonlocal} and $\vec{k}_{1}$ the three-momentum of one outgoing particle,
$M_{K}$ the kaon mass, and $M_{\pi}$ the pion mass. (For details and
phenomenology of the spectral function, see Refs. \cite{lupo}).

The propagator of $K_{0}^{\ast}$ is given by $\Delta_{K_{0}^{\ast}}%
(m^{2})=\left[  m^{2}-M_{0}^{2}+\Pi(m^{2})+i\varepsilon\right]  ^{-1}$,
$M_{0}$ being the bare mass of $K_{0}^{\ast}(1430)$ and $\Pi(m^{2})$ the
one-loop contribution. The spectral function $d_{K_{0}^{\ast}}(m)=\frac
{2m}{\pi}|{\operatorname{Im}}\Delta_{K_{0}^{\ast}}(p^{2}=m^{2})|$ is the mass
probability density (its integral is normalized to unity). Typically, for the
\textquotedblleft Breit-Wigner\textquotedblright\ value $M_{BW}$ determined as
$M_{BW}^{2}-M_{0}^{2}+\operatorname{Re}\Pi(M_{BW}^{2})=0$ the spectral
function has a peak's width $\Gamma_{BW}=\operatorname{Im}\Pi(M_{BW})/M_{BW}.$
A useful approximation, valid if the width is sufficiently small, is the
relativistic Breit-Wigner expression:%

\begin{equation}
d_{K_{0}^{\ast}}(m)\approx d_{K_{0}^{\ast}}^{BW}(m)=N\left[  \left(
m^{2}-M_{BW}^{2}\right)  ^{2}+M_{BW}^{2}\Gamma_{BW}^{2}\right]  ^{-1}\text{ .}%
\end{equation}
Under this approximation, there is only one pole in the complex plane at
$m^{2}\simeq M_{BW}^{2}-iM_{BW}\Gamma_{BW}$ (hence, $m\simeq M_{BW}%
-i\Gamma_{BW}/2$). But, when a resonance is broad, these approximations are
not anymore valid.

We now turn to $\pi K$ scattering. Within our framework, the pion-kaon phase
shift is given by \cite{milena}:
\begin{equation}
\delta_{\pi K,swave}(m)=\delta_{(I=1/2,J=0)}(m)=\frac{1}{2}\arccos\left[
1-\pi\Gamma_{K_{0}^{\ast}}(m)d_{K_{0}^{\ast}}(m)\right]  \ \text{,}%
\label{phaseshift}%
\end{equation}
where $\delta_{(I,J)}(m)$ is the general phase shift for a given isospin $I$
and total spin $J$. The amplitude of the process and the phase-shift are
linked by $a_{(I,J)}=\left(  e^{i\delta_{(I,J)}(m)}-1\right)  /(2i)$. The
parameters $(a,b,M_{0,}\Lambda)$ entering in\ Eq. (\ref{lagk}) were determined
via a fit to $\pi K$ phase-shift data \cite{pionkaonexp}, see Ref.
\cite{milena} for details. A very good description of data is achieved. A
study of the complex plane shows an interesting fact: besides the pole
corresponding to the well-known $K_{0}^{\ast}(1430)$ state $(1.413\pm
0.002)-i(0.127\pm0.003)$ GeV, there is a second pole which correspond to
$K_{0}^{\ast}(700)$:
\begin{equation}
(0.746\pm0.019)-i(0.262\pm0.014)\text{ GeV.}\label{ourpolek}%
\end{equation}
The numerical value is compatible with the PDG value of Eq. (\ref{pdgpolek}).
A large-$N_{c}$ study confirms that, while the first pole tends to the real
axis (and hence is a $\bar{q}q$ state), the second one moves away from it, as
it is expected for a dynamically generated state.

In conclusion, the simple model of Eq. (\ref{lagk}) is able to describe $\pi
K$ scattering data and naturally gives rise to the pole of $K_{0}^{\ast}(700)$
as a companion pole of the predominantly quark-antiquark resonance
$K_{0}^{\ast}(1430).$

\section{The light $\kappa$ at nonzero temperature}

The partition function of an hadronic gas can be expressed as the sum of the
contributions of stable particles and their mutual interactions:
\begin{equation}
\ln Z=\ln Z_{\text{pions}}+\ln Z_{\text{kaons}}+...+\ln Z^{int}\text{, }\ln
Z^{int}=\text{ }%
{\displaystyle\sum\limits_{I,J}}
\ln Z_{(I,J)}\text{ .}%
\end{equation}
The first term $\ln Z_{\text{pions}}=3F_{1}(m_{\pi})$ refers to pions and $\ln
Z_{\text{kaons}}=4F_{1}(m)$ to kaons, where $F_{1}(m)=\int\frac{\mathrm{d}%
^{3}\mathrm{p}}{(2\pi)^{3}}\ln\left[  1-e^{-\sqrt{\vec{p}^{2}+m_{\pi}^{2}}%
/T}\right]  $ is the contribution of a free particle with mass $m$. The term
$\ln Z_{IJ}$ refers to the contribution of the interactions in the $(I,J)$
channel \cite{dashen}:%

\begin{equation}
\ln Z_{(I,J)}=(2I+1)(2J+1)\int_{0}^{\infty}\frac{\mathrm{dm}}{\pi}%
\frac{d\delta_{(I,J)}(m)}{dm}F_{1}(m)\text{ .} \label{lnzij}%
\end{equation}
When in a certain channel a narrow resonance is present, one finds its
standard contribution. For instance, for $I=J=1$ the $\rho$ meson is produced.
In the nonrelativistic BW-limit $\frac{1}{\pi}\frac{d\delta_{(1,1)}(m)}%
{dm}\simeq\frac{\Gamma_{\rho}}{2\pi}\left[  (m-M_{\rho})^{2}+\Gamma_{\rho}%
^{2}/4\right]  ^{-1}$. (Moreover, for $\Gamma_{\rho}\rightarrow0$,
$\delta(m-M_{\rho})$ emerges: the contribution of a stable $\rho$ is obtained.).

However, Eq. (\ref{lnzij}) is very general and can describe also broad
resonances as well as non-resonant channels, such as repulsive ones. This is
important for the $\kappa$. In the resonant $I=1/2,J=0$ channel in which the
$\kappa$ is formed, one has (upon integrating up to $1$ GeV) $\ln
Z_{(1/2,0)}=\int_{0}^{1\text{ GeV}}\frac{2\mathrm{dm}}{\pi}\frac
{d\delta_{(1/2,0)}(m)}{dm}F_{1}(m)$. This is sizable. However, one should also
consider the repulsion in the $I=3/2,$ $J=0$ channel.\ Remarkably, the sum
\begin{equation}
\ln Z_{(1/2,0)}+\ln Z_{(3/2,0)}=\int_{0}^{1\text{ GeV}}\mathrm{dm}\left(
\frac{2}{\pi}\frac{d\delta_{(1/2,0)}(m)}{dm}+\frac{4}{\pi}\frac{d\delta
_{(3/2,0)}(m)}{dm}\right)  F_{1}(m)
\end{equation}
is \textit{small}. Namely, while $\frac{d\delta_{(1/2,0)}(m)}{dm}>0$
(attraction), $\frac{4}{\pi}\frac{d\delta_{(3/2,0)}(m)}{dm}<0$ (repulsion).
Note: $\frac{1}{\pi}\frac{d\delta_{(1/2.0)}(m)}{dm}\neq d_{K_{0}^{\ast}}(m)$.
(This would be true only in the BW limit). In conclusion, the light $\kappa$
can be safely neglected in the construction of thermal hadronic models.

\section{Conclusions}

We have described the emergence of the state $\kappa\equiv K_{0}^{\ast}(700)$
as a companion pole of $K_{0}^{\ast}(1430)$ by using an effective hadronic
model \cite{milena}.\ The numerical value of the pole (\ref{ourpolek}) is in
agreement with the present PDG estimate of Eq. (\ref{pdgpolek}). On the other
hand, contrary to the naive expectations, the light $\kappa$ is not relevant
in a thermal hadronic gas. Namely, its influence on thermodynamical properties
is cancelled by a repulsion in the $I=3/2$ channel. Either one includes both
the light $\kappa$ and the repulsion, or -even easier- neglects both of them.

\bigskip

\textbf{Acknowledgements:} The author thanks M. Piotrowska, T.\ Wolkanowski,
W. Broniowski, V.\ Begun for cooperations. Financial support from the Polish
National Science Centre (NCN) through the OPUS project no. 2015/17/B/ST2/01625
is acknowledged.


\begin{thebibliography}{99}                                                                                               %


\bibitem {pdg}M. Tanabashi et al. (Particle Data Group), Phys. Rev. D
\textbf{98}, 030001 (2018).

\bibitem {pdg2016}C. Patrignani et al. (Particle Data Group), Chin. Phys. C
\textbf{40} 100001 (2016).

\bibitem {kpoles}
S.~Ishida, M.~Ishida, T.~Ishida, K.~Takamatsu and T.~Tsuru,
Prog.\ Theor.\ Phys.\ \textbf{98} (1997) 621 [hep-ph/9705437].
D.~Black, A.~H.~Fariborz, F.~Sannino and J.~Schechter,
Phys.\ Rev.\ D \textbf{58} (1998) 054012.
P.~C.~Magalhaes \textit{et al.},
Phys.\ Rev.\ D \textbf{84} (2011) 094001 [arXiv:1105.5120 [hep-ph]].
S.~Descotes-Genon and B.~Moussallam,
Eur.\ Phys.\ J.\ C \textbf{48} (2006) 553 [hep-ph/0607133].
J.~R.~Pelaez,
Phys.\ Rev.\ Lett.\ \textbf{92} (2004) 102001 [hep-ph/0309292].
J.~R.~Pelaez and A.~Rodas,
Eur.\ Phys.\ J.\ C \textbf{77} (2017) no.6, 431
[arXiv:1703.07661 [hep-ph]].
P.~Buettiker, S.~Descotes-Genon and B.~Moussallam,
Eur.\ Phys.\ J.\ C \textbf{33}, 409 (2004) [hep-ph/0310283].
J.~Sa Borges, J.~Soares Barbosa and V.~Oguri,
Phys.\ Lett.\ B \textbf{412} (1997) 389.
H.~Q.~Zheng, Z.~Y.~Zhou, G.~Y.~Qin, Z.~Xiao, J.~J.~Wang and N.~Wu,
Nucl.\ Phys.\ A \textbf{733} (2004) 235 [hep-ph/0310293].
Z.~Y.~Zhou and H.~Q.~Zheng,
Nucl.\ Phys.\ A \textbf{775} (2006) 212 [hep-ph/0603062].
A.~H.~Fariborz, E.~Pourjafarabadi, S.~Zarepour and S.~M.~Zerbarjad,
Phys.\ Rev.\ D \textbf{92} (2015) 113002
[arXiv:1511.01623 [hep-ph]].
S.~Descotes-Genon and B.~Moussallam,
Eur.\ Phys.\ J.\ C \textbf{48} (2006) 553 [hep-ph/0607133].
M.~Ablikim \textit{et al.} [BES Collaboration],
Phys.\ Lett.\ B \textbf{698} (2011) 183 [arXiv:1008.4489 [hep-ex]].


\bibitem {pelaezrev}
J.~R.~Pelaez,
Phys.\ Rept.\ \textbf{658} (2016) 1
[arXiv:1510.00653 [hep-ph]].


\bibitem {tq}R.~L.~Jaffe,
Phys.\ Rev.\ D \textbf{15} (1977) 267.
R.~L.~Jaffe,
Phys.\ Rev.\ D \textbf{15} (1977) 281.
R.~L.~Jaffe,
Phys.\ Rept.\ \textbf{409} (2005) 1 [Nucl.\ Phys.\ Proc.\ Suppl.\ \textbf{142}
(2005) 343] [arXiv:hep-ph/0409065].
L.~Maiani, F.~Piccinini, A.~D.~Polosa and V.~Riquer,
Phys.\ Rev.\ Lett.\ \textbf{93}, 212002 (2004) [arXiv:hep-ph/0407017].
F.~Giacosa,
Phys.\ Rev.\ D \textbf{74} (2006) 014028 [arXiv:hep-ph/0605191]. F.~Giacosa,
Phys.\ Rev.\ D \textbf{75} (2007) 054007 [arXiv:hep-ph/0611388].
A.~H.~Fariborz, R.~Jora and J.~Schechter,
Phys.\ Rev.\ D \textbf{72} (2005) 034001 [hep-ph/0506170].
M.~Napsuciale and S.~Rodriguez,
Phys.\ Rev.\ D \textbf{70} (2004) 094043 [hep-ph/0407037].

\bibitem {lowscalars}E.~van Beveren \textit{et al.},
Z.\ Phys.\ C \textbf{30}, 615 (1986) [arXiv:0710.4067 [hep-ph]]. E.~van
Beveren, D.~V.~Bugg, F.~Kleefeld and G.~Rupp,
Phys.\ Lett.\ B \textbf{641}, 265 (2006) [arXiv:hep-ph/0606022].
J.~R.~Pelaez,
Phys.\ Rev.\ Lett.\ \textbf{92}, 102001 (2004) [arXiv:hep-ph/0309292].
J.~A.~Oller and E.~Oset,
Nucl.\ Phys.\ A \textbf{620}, 438 (1997) [arXiv:hep-ph/9702314].
J.~A.~Oller, E.~Oset and J.~R.~Pelaez,
Phys.\ Rev.\ D \textbf{59} (1999) 074001 [arXiv:hep-ph/9804209]. J. A. Oller,
E. Oset and J. R. Pel\'{a}ez, Phys. Rev. Lett\emph{.} \textbf{80}, 3452-3455 (1998).

\bibitem {isgur}
S.~Godfrey and N.~Isgur,
\ Phys.\ Rev.\ D \textbf{32} (1985) 189.


\bibitem {elsm}D. Parganlija, P. Kovacs, G. Wolf, F. Giacosa and D. H.
Rischke, Phys. Rev. \textbf{D87}, 014011 (2012). S.~Janowski, F.~Giacosa and
D.~H.~Rischke,
Phys.\ Rev.\ D \textbf{90} (2014) 11, 114005 [arXiv:1408.4921 [hep-ph]].


\bibitem {milena}
T.~Wolkanowski, M.~So\l tysiak and F.~Giacosa,
Nucl.\ Phys.\ B \textbf{909} (2016) 418
[arXiv:1512.01071 [hep-ph]].


\bibitem {a0}
T.~Wolkanowski, F.~Giacosa and D.~H.~Rischke,
Phys.\ Rev.\ D \textbf{93} (2016) no.1, 014002
[arXiv:1508.00372 [hep-ph]].
M.~Boglione and M.~R.~Pennington,
Phys.\ Rev.\ D \textbf{65} (2002) 114010
[hep-ph/0203149].


\bibitem {andronic}
A.~Andronic, P.~Braun-Munzinger and J.~Stachel,
Nucl.\ Phys.\ A \textbf{772} (2006) 167 [nucl-th/0511071].
A.~Andronic, P.~Braun-Munzinger and J.~Stachel,
Phys.\ Lett.\ B \textbf{673} (2009) 142 [Phys.\ Lett.\ B \textbf{678} (2009)
516] [arXiv:0812.1186 [nucl-th]].
P.~Alba \textit{et al},
Phys.\ Lett.\ B \textbf{738} (2014) 305 [arXiv:1403.4903 [hep-ph]].
G.~Torrieriet \textit{et al}.,
Comput.\ Phys.\ Commun.\ \textbf{167} (2005) 229 [nucl-th/0404083].


\bibitem {bronio}
W.~Broniowski, F.~Giacosa and V.~Begun,
Phys.\ Rev.\ C \textbf{92} (2015) no.3, 034905
[arXiv:1506.01260 [nucl-th]].


\bibitem {dashen}
R.~Dashen, S.~K.~Ma and H.~J.~Bernstein,
Phys.\ Rev.\ \textbf{187} (1969) 345.
R.~F.~Dashen and R.~Rajaraman,
Phys.\ Rev.\ D \textbf{10} (1974) 694.
W.~Weinhold, B.~L.~Friman and W.~Noerenberg,
Acta Phys.\ Polon.\ B \textbf{27} (1996) 3249.
W.~Weinhold, B.~Friman and W.~Norenberg,
Phys.\ Lett.\ B \textbf{433} (1998) 236 [nucl-th/9710014].


\bibitem {hiller}
W.~Broniowski, W.~Florkowski and B.~Hiller,
Phys.\ Rev.\ C \textbf{68} (2003) 034911 [nucl-th/0306034].


\bibitem {pok}
P.~M.~Lo,
Eur.\ Phys.\ J.\ C \textbf{77} (2017) no.8, 533
[arXiv:1707.04490 [hep-ph]].
P.~M.~Lo, B.~Friman, M.~Marczenko, K.~Redlich and C.~Sasaki,
Phys.\ Rev.\ C \textbf{96} (2017) no.1, 015207
[arXiv:1703.00306 [nucl-th]].


\bibitem {nonlocal}J.~Terning,
Phys.\ Rev.\ D \textbf{44} (1991) 887.
A.~Faessler, T.~Gutsche, M.~A.~Ivanov, V.~E.~Lyubovitskij and P.~Wang,
Phys.\ Rev.\ D \textbf{68} (2003) 014011 [arXiv:hep-ph/0304031].
F. Giacosa, T. Gutsche and A. Faessler,
\ Phys. Rev. C 71, 025202 (2005) [arXiv:hep-ph/0408085].

\bibitem {lupo}
F.~Giacosa and G.~Pagliara,
Phys.\ Rev.\ C \textbf{76} (2007) 065204
[arXiv:0707.3594 [hep-ph]].
S.~Coito and F.~Giacosa,
arXiv:1712.00969 [hep-ph].


\bibitem {pionkaonexp}
D.~Aston \textit{et al.},
Nucl.\ Phys.\ B \textbf{296} (1988) 493.

\end{thebibliography}
\end{document}